\documentclass[prl,twocolumn,showpacs,preprintnumbers,amsmath,amssymb]{revtex4}
\usepackage{graphicx}
\usepackage{dcolumn} 
\usepackage{bm}      
\newcommand{\be}{\begin{equation}}
\newcommand{\ee}{\end{equation}}
\newcommand{\bea}{\begin{eqnarray}}
\newcommand{\eea}{\end{eqnarray}}
\newcommand{\mm}{\mathrm}

\newcommand{\mc}{\mathcal}

\newcommand{\bi}{\begin{itemize}}
\newcommand{\ei}{\end{itemize}}

\begin{document}
  \title{Random Field Ising Model In and Out of Equilibrium}
  \author{Yang Liu}
  \author{Karin A.\ Dahmen}
  \affiliation{Department of Physics, University of Illinois at
    Urbana-Champaign, Urbana, IL 61801, USA}
  \date{\today}
  \begin{abstract} 
    We present numerical studies of zero-temperature Gaussian random-field Ising
    model (zt-GRFIM) in both equilibrium and non-equilibrium. We compare the
    no-passing rule, mean-field exponents and universal quantities
    in 3D (avalanche critical exponents, fractal dimensions, scaling functions
    and anisotropy measures) for the equilibrium and non-equilibrium
    disorder-induced phase transitions. We show compelling evidence that the
    two transitions belong to the same universality class.   
  \end{abstract}
  \pacs{02.60.Pn, 75.10.Nr, 75.60.Ej, 64.60.Fr}
  \maketitle

  As a prototypical model for magnets with quenched disorder, the
  random-field Ising Model (RFIM) has been intensively studied
  during the last thirty years~\cite{Nattermann-98}. Nevertheless, some
  theoretically and experimentally important questions are still not well
  answered. For example, it is still controversial whether the equilibrium
  and non-equilibrium disorder-induced phase transitions of the
  zero-temperature RFIM belong to the same universality class.

  The RFIM is defined by the Hamiltonian  \be {\cal H} = - \!\sum_{{<}i,j{>}} J \, s_i
  s_j - \sum_i \, (H + h_i) \,  s_i\ee where the spins $s_i = \pm 1$ sit on a
  $D$-dimensional hypercubic lattice with periodic boundary conditions. The
  spins interact ferromagnetically with their nearest neighbors with strength
  $J$ and experience a uniform external field $H$ and a local field $h_i$. To
  model quenched disorders, the local fields $h_i$ are randomly chosen from a Gaussian 
  distribution with mean zero and variance $R$. $R$ is often called the
  disorder parameter or just disorder.    

  In equilibrium, it is well known that for $D\ge3$ there is a
  continuous phase transition between the ferromagnetic and paramagnetic
  phases at finite temperatures and disorders~\cite{Nattermann-98}. Critical
  behavior of this transition is controlled by a stable zero-temperature fixed
  point~\cite{Villain-84, Fisher-86}. Therefore, we can stay at $T=0$ and
  study the phase transition undergone by the ground state (GS) as $R$ is
  tuned to the critical value $R_\mm{c}$, i.e. the disorder-induced phase transition
  (DIPT), to obtain the equilibrium properties at finite
  temperatures~\cite{Ogielski-86-b}.  As for the GS 
  problem of the RFIM, it can be mapped onto the min-cut/max-flow problem in
  combinatorial optimization and then solved via the so-called push-relabel
  algorithm~\cite{Cherkassky-97}. In non-equilibrium, the DIPT was first 
  numerically observed by Sethna \emph{et al.} in the hysteretic behavior at
  $T=0$ and $D\ge3$~\cite{Sethna-93}. A local metastable 
  dynamics was introduced there: As $H$ is slowly increased from $-\infty$ to
  $\infty$ and decreased back to $-\infty$, each spin flips deterministically
  when its effective local field $h^{\rm eff}_i = J \sum_{j} s_j + h_i + H $
  changes sign. It is found that there is a critical point ($R_\mm{c}$, $H_\mm{c}$)
  which separates macroscopically smooth saturation hysteresis loops in the
  magnetization $M(H)$ (for $R>R_\mm{c}$) from saturation loops with a
  macroscopic jump or burst (for $R<R_\mm{c}$). Here, $H_\mm{c}$ is the
  non-universal magnetic field value at which the magnetization curve has
  infinite slope. (Of course, in equilibrium $H_\mm{c}=0$ due to symmetry.)
  This non-equilibrium DIPT has also been studied analytically~\cite{Karin-96}
  and experimentally~\cite{Berger-00}.

  Comparing the equilibrium and non-equilibrium DIPTs is very interesting. In
  mean field theory (MFT), they have the same thermodynamic critical exponents
  and the same exponent relations~\cite{Schneider-77,
  Sethna-93}. Renormalization group (RG) calculation shows that the
  $6-\epsilon$ expansion for the non-equilibrium critical exponents maps to
  all orders in $\epsilon$ onto the equilibrium ones~\cite{Karin-96}, though
  the RG description of the equilibrium RFIM has been controversial for
  decades~\cite{Feldman-02}. In 3D and even 4D, 
  numerical values of the critical exponents of the two DIPTs seem to match within
  the error bars~\cite{Maritan-94,Perkovic-99,Hartmann-02}.

  Recently, Vives \emph{et al.} suggested that the two DIPTs belong to the
  same universality class by conjecturing the extrapolation result of a RG
  type argument~\cite{Vives-04}. Meanwhile, Colaiori \emph{et al.} numerically
  compared the equilibrium DIPT, i.e. the DIPT of the GS, to that of the
  demagnetized state (DS), considering the DS as a non-equilibrium hysteretic
  counterpart of the GS often used in experiments and
  applications~\cite{Colaiori-04}. Here, the DS 
  is obtained by applying an external oscillating field with slowly decreasing
  amplitude to the non-equilibrium system. The system will then be taken
  through a series of subloops and the line connecting the tips of those 
  subloops is known as the demagnetization curve. Colaiori \emph{et al.} compare the
  scaling behavior of the magnetization $M$ for the DS and the GS near their
  respective $R_\mm{c}$ and at $H_\mm{c}$. ($H_\mm{c}=0$ for both cases.)
  Doing finite-size scaling with the known thermodynamic critical exponents,
  they present evidence that the DIPT of the DS and that of the GS are in the same
  universality class, in both 3D and the Bethe lattice. On the other hand,
  Carpenter \emph{et al.} found a related DIPT for the demagnetization curve,
  which displays similar critical behavior as that of the saturation
  loop~\cite{Carpenter-03}.

  Despite those evidences in favor of universality, the original question is
  still not fully answered. We notice that some important critical exponents
  and universal quantities of the DIPT of the GS and of the saturation loop
  have never been compared. Also, we notice that by comparing universal
  scaling functions rather than just critical exponents, we are comparing an
  infinite amount of more information than was done previously. (1) \emph{We
  compare the avalanche exponents and scaling functions associated with the
  avalanche size distribution.} Here, ``avalanche'' refers to the flip of
  neighboring spins during the magnetization process, corresponding to a jump
  in the magnetization curve $M(H)$. The number of spins participating in an
  avalanche is called its size $S$. Since in non-equilibrium the avalanche
  exponents and the associated scaling function have been well studied,
  comparing them with the corresponding equilibrium ones constitutes a
  particularly strong test for universality.  (2) \emph{We compare the spatial
  structure of avalanches and clusters near the critical disorder.} Here,
  clusters are connected regions of flipped spins, formed by the aggregation
  of avalanches. In non-equilibrium, it is known that near the critical
  disorder the spatial structure of avalanches is visually interesting:
  fractal and anisotropic~\cite{Sethna-01}. In equilibrium it also has been
  found that near the critical disorder clusters have fractal
  surfaces~\cite{Middleton-02}. Up to now, the comparison of spatial
  structures of avalanches (or clusters) in equilibrium and non-equilibrium
  has never been done. This would be another independent test of the
  universality. We think the whole idea of looking at avalanches and clusters
  is quite neat: Not only are we directly testing more sensitive features of
  the problem, but we are giving insight into why the two DIPTs could be
  similar: the equilibrium and non-equilibrium systems could have similar
  avalanches and clusters during the magnetization process.

  In equilibrium, the magnetization process can be simulated with the
  efficient algorithm reported in Ref.~\onlinecite{Hartmann-98,Vives-00}. This
  algorithm is essentially based on the fact that the GS energy has a
  convexity property which allows for estimates of the fields where the
  magnetization jumps (called ``avalanches'' occur). In non-equilibrium, three
  different algorithms to simulate the magnetization process (hysteresis
  loops) are described in Ref.~\onlinecite{Kuntz-99}. All these algorithms use
  the adiabatic single-spin-flip dynamics introduced in
  Ref.~\onlinecite{Sethna-93}. In this work, we have studied the magnetization
  processes in both equilibrium and non-equilibrium for system sizes ranging
  from $L^3=32^3$ to $192^3$. All the measured properties are averaged over a
  large number of realizations of the random-field configuration. Typical
  averages are performed over a number of realizations that ranges between
  $10^4$ for $L=32$ and $50$ for $L=192$.

  Before we present any numerical results in 3D, we show two additional
  similarities beyond the 3D simulation. (1) \emph{The avalanche critical 
  exponents in MFT must be the same for the two 
  DIPTs.} We start the proof by noticing that in non-equilibrium the hard spin MFT
  magnetization curve has no hysteresis for $R \ge R_\mm{c}$~\cite{Karin-96}. Since there is
  only one $M(H)$ solution for $R \ge R_\mm{c}$ in MFT, it must be the
  non-equilibrium and the equilibrium solution at the same time. In MFT, every
  spin couples to $M(H)$. Since $M(H)$ is unique, this implies that as $H$ is
  increased there is a unique series of local-field configurations and
  therefore a unique series of states. This means that in MFT for $ R \ge
  R_\mm{c}$ the avalanches in equilibrium must be the same as the avalanches
  in non-equilibrium. So the MFT avalanche exponents must be the same in both
  DIPTs.  (2)\emph{ Middleton's
  no-passing rule}~\cite{Middleton-92}: One defines the natural partial ordering
  of two states: a state $C= \{s_1,\cdots,s_N\} \ge
  \tilde{C}=\{\tilde{s}_1,\cdots,\tilde{s}_N\}$ if $s_i \ge \tilde{s}_i$ for each site $i$ of the
  system. Let a system $C(t)$ be evolved under the fields $H(t)$
  and similarly $\tilde{C}(t)$ evolved under $\tilde{H}(t)$. Suppose the fields
  $H(t)\ge \tilde{H}(t)$ and the initial states satisfy
  $C(0)\ge\tilde{C}(0)$, then the no-passing rule guarantees the
  partial ordering will be preserved, i.e. $C(t)\ge\tilde{C}(t)$ at
  all times later $t>0$. For the magnetization process, this is equivalent to the
  absence of reverse spin flips as $H$ is swept from $-\infty$ to $\infty$. 
  In non-equilibrium, this rule has been proven and applied to explain the
  return-point memory~\cite{Sethna-93}. In equilibrium, the main idea of the proof
  follows. For any state $C_2$ at field $H_2$ which evolves from the GS $C_1$
  at field $H_1$ ($H_1<H_2$) with reverse avalanches, we can always find a
  corresponding state $\tilde{C}$ which evolves from $C_1$ without any reverse
  avalanches and has lower energy than $C_2$ at field $H_2$.  So as $H$ is
  increased, the GS evolves without any reverse spin flips. Since flipped
  spins need not be considered any more in the GS calculation for all higher
  fields, the algorithm will be accelerated dramatically~\cite{Vives-00}. For
  details, see Ref.~\onlinecite{Liu-06}.

  \begin{figure}[t]
    \includegraphics[width=0.48\textwidth]{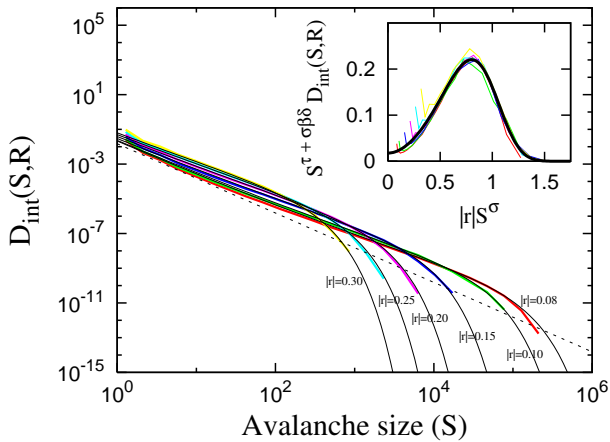}
   \caption{\label{fig:Dint}(Color online) Integrated equilibrium avalanche
    size distribution curves in 3D for $64^3$ spins and different
    disorders. Those curves are averaged up to 500 initial random-field
    configurations. The inset shows the scaling collapse of the integrated
    avalanche size distribution, using $(\tau+\sigma\beta\delta)^\mm{eq}=2.00$
    and $\sigma^\mm{eq}=0.23$. (Even with
    $(\tau+\sigma\beta\delta)^\mm{neq}=2.03$ and $\sigma^\mm{neq}=0.24$, the
    collapse still looks good.) The thick black curve through the collapse is
    the non-equilibrium universal scaling function
    $\bar{\mc{D}}_{-}^{\mm{int}}(X)$ (see text). In the main panel, the
    equilibrium distribution curves obtained from the non-equilibrium scaling
    function are plotted (thin solid lines) alongside the raw data (thick
    solid lines). The straight dashed line is the expected asymptotic
    power-law behavior: $S^{-2.00}$, which does not agree with the measured
    slope of the raw data quoted in Ref.~\onlinecite{Vives-02} due to the
    ``bump'' in the scaling function.} 
  \end{figure}

  Now, we report new 3D simulation results which present evidence of universality for the
  two DIPTs. First, we extract the avalanche exponents from the field integrated avalanche 
  size distribution $D_\mm{int}(S,R)$ associated with the equilibrium
  magnetization curve, see Fig.~\ref{fig:Dint}.  In both
  equilibrium and non-equilibrium, the scaling form of $D_\mm{int}(S,R)$ can
  be written as \be D_\mm{int}(S,R) \sim  S^{-(\tau+\sigma\beta\delta)} \ 
  \bar{\mc{D}}_{\pm}^{\mm{int}} (S^{\sigma}|r|) \label{Dint-SR} \ee with
  $r=(R_\mm{c}-R)/R$. Note that $R_\mm{c}$ is non-universal. For Gaussian
  disorders in 3D, $R_\mm{c}^\mm{eq}=2.270\pm0.004$~\cite{Middleton-02} and
  $R_\mm{c}^\mm{neq}=2.16\pm0.03$~\cite{Perkovic-99}.  In non-equilibrium, the
  quantity $D_\mm{int}(S,R)$ has been studied extensively, where
  $(\tau+\sigma\beta\delta)^{\mm{neq}}=2.03 \pm 0.03$ and
  $\sigma^{\mm{neq}}=0.24\pm0.02$ were obtained from scaling collapses and
  linear extrapolation to $R_\mm{c}$~\cite{Sethna-93,Perkovic-99}. In
  equilibrium, using the same method, we have
  $(\tau+\sigma\beta\delta)^{\mm{eq}}=2.00 \pm 0.01$ and
  $\sigma^{\mm{eq}}=0.23\pm0.01$. Both $\sigma^{\mm{eq}}$ and
  $(\tau+\sigma\beta\delta)^{\mm{eq}}$ match their non-equilibrium
  values. Plotting the non-equilibrium universal scaling
  function~\cite{Perkovic-99}: $\bar{\mc{D}}_{-}^{\mm{int}} (X) = e^{-0.789
  X^{1/\sigma}} (0.021 + 0.002 X + 0.531 X^2 - 0.266 X^3 + 0.261X^4 )$ on
  top of the equilibrium collapse, we find an excellent match, up to the
  overall horizontal and vertical scaling factors, see inset of
  Fig.~\ref{fig:Dint}. According to this scaling function, we plot the
  distribution curves on top of the original data, we find excellent fits for
  all disorders. The match in both critical exponents and scaling functions
  strongly indicate that the two DIPTs belong to the same universality class.

  Second, we consider the spatial structure of avalanches and clusters at
  $R_\mm{c}$ in both equilibrium and non-equilibrium. Avalanches are
  integrated over the field $H$ while clusters are chosen from states near the
  critical field $H_\mm{c}$. ($H_\mm{c}^\mm{eq}=0$ and
  $H_\mm{c}^\mm{neq}=1.435\pm0.004$ at $R=R_\mm{c}$ for Gaussian disorder in
  3D~\cite{Perkovic-99}.)  The spatial structure can be quantitatively described by fractal
  dimensions and anisotropy measures. We first compute the fractal dimensions
  $d_\mm{f}$ of the size (or mass) $S$, enclosed volume $v$ and outermost
  surface $a$ of avalanches and clusters. Note that the difference between $v$
  and $S$ is due to possible ``holes'' inside avalanches (or
  clusters)~\cite{Middleton-02}. For finite systems, the natural finite-size 
  scaling hypothesis reads $f(l;R,L)=L^{d_\mm{f}} \hat{f}(rL^{1/\nu}, l/L)
  \label{eq:f(l,R,L)}.$ Here, $f$ could be $S$, $v$ or $a$ of the avalanche
  (or cluster) with linear size $l$ in a system of linear size $L$. $\nu$ is
  the critical exponent of the correlation length $\xi$ and $\hat{f}$ is a
  universal scaling function. This hypothesis enables us to do the
  scaling collapse at $R_\mm{c}$ for different system sizes. Extrapolation
  values ($L \to \infty $) are quoted in Table~\ref{tab:exponent}. Here, the
  error bars include both statistical and systematic errors. We find that
  $d_\mm{S} = d_\mm{v}$ for all the cases, which indicates the ``holes'' would
  be ignorable in the thermodynamic limit. Moreover, considering the
  systematic errors could be even larger than the ones listed here, we
  conclude that the fractal dimensions ($d_\mm{S}$, $d_\mm{v}$ or $d_\mm{a}$)
  of avalanches (or clusters) in equilibrium and non-equilibrium are very
  close. For the anisotropy measures, as done in the percolation 
  and polymer systems~\cite{Jagodzinski-92}, we use the radius of gyration
  tensor $Q$: $Q_{\alpha \beta}=\frac{1}{2 S^2}
  \sum_{i,j=1}^{S}[X_{i,\alpha}-X_{j,\alpha}] 
  [X_{i,\beta}-X_{j,\beta}]$ to characterize the shape of a given conformation
  of $S$ points in a $D$-dimensional hypercubic lattice, i.e. an avalanche or
  a cluster of size $S$. Here $X_{i,\alpha}$ is the $\alpha$-coordinate 
  of the $i$-th point with $\alpha=1,...,D$. All the anisotropy measures are
  related to $Q$'s $D$ eigenvalues: $\lambda_{\alpha}$ with
  $\lambda_1 \ge \lambda_2 \ge \cdots \ge \lambda_D $ and
  $\bar{\lambda}=(\sum_{\alpha} \lambda_{\alpha})/D$. (1) \emph{Anisotropy}
  $A_1\equiv\lambda_D/\lambda_1$ and
  $A_2\equiv\lambda_\mm{D-1}/\lambda_1$, which are the simplest anisotropy
  measures in 3D. (2) \emph{Asphericity}
  $\Delta_D=\frac{1}{D(D-1)}\sum^D_{\alpha=1}\frac{(\lambda_{\alpha}-\bar{\lambda})^2}{\bar{\lambda}^2}$,
  which characterizes the shape's overall deviation from spherical
  symmetry. (3) \emph{Prolateness} $S_D\equiv
  (\lambda_1-\bar{\lambda})(\lambda_2-\bar{\lambda})(\lambda_3-\bar{\lambda})/(2\bar{\lambda}^3)$, which distinguishes prolate (positive $S_D$) from oblate shapes
  (negative $S_D$) in 3D.  Asymptotic values of those anisotropy measures in
  the large size limit can be obtained from the largest avalanches (or
  clusters) that are not affected by finite-size effects. Results are shown in
  Table~\ref{tab:exponent}. It is interesting to mention that avalanches 
  (or clusters) are prolate in both equilibrium and non-equilibrium. More
  importantly, we find that the asymptotic values of all the anisotropy
  measures of avalanches (or clusters) in equilibrium and non-equilibrium are
  very close.

  \begin{table}[t]
    \caption{\label{tab:exponent} Fractal dimensions and anisotropy measures
      obtained from numerical simulations in 3D for both equilibrium and
      non-equilibrium zt-GRFIM.}
    \begin{ruledtabular}
      \begin{tabular}{ccccc}
	&\multicolumn{2}{c}{non-equilibrium}&\multicolumn{2}{c}{equilibrium}\\
	Quantities    & Avalanches & Clusters & Avalanches & Clusters  \\ \hline
	\hline
	$d_{\mm{S}}$ & $2.78\pm0.05$ & $2.76\pm0.04$ & $2.77\pm0.09$ & $2.78\pm0.05$\\
	$d_{\mm{v}}$ & $2.78\pm0.05$ & $2.76\pm0.04$ & $2.77\pm0.09$ & $2.78\pm0.05$\\
	$d_{\mm{a}}$ & $2.33\pm0.04$ & $2.18\pm0.04$ & $2.16\pm0.05$ & $2.11\pm0.03$\\

	\hline
	$A_1$      & $0.29\pm0.01$ & $0.25\pm0.01$ & $0.30\pm0.02$ & $0.28\pm0.01$   \\
	$A_2$      & $0.50\pm0.02$ & $0.45\pm0.02$ & $0.50\pm0.02$ & $0.48\pm0.02$   \\
	$\Delta_D$ & $0.16\pm0.01$ & $0.21\pm0.02$ & $0.16\pm0.02$ & $0.18\pm0.02$ \\
	$S_D$      & $0.06\pm0.01$ & $0.09\pm0.01$ & $0.06\pm0.01$ & $0.07\pm0.01$  \\
      \end{tabular}
    \end{ruledtabular}
  \end{table}

\begin{figure}[t]
    \includegraphics[width=0.45\textwidth]{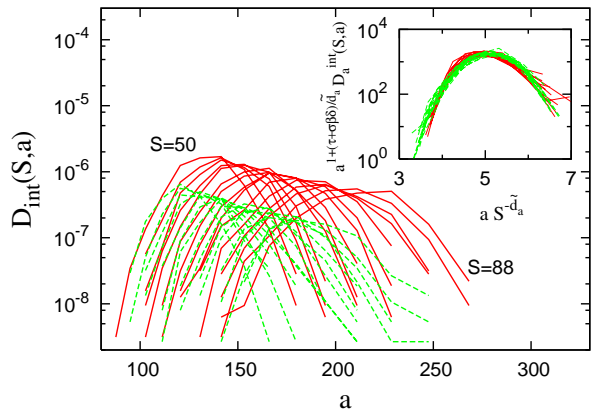}
    \caption{\label{fig:Dint_a-S} (Color online) Avalanche surface area
      distribution curves in 3D at $R_\mm{c}$, for avalanche size bins from 50
      to 88 (from upper left to lower right corner). The system size is
      $192^3$ for non-equilibrium (solid lines, averaged over 45 initial
      random-field configurations) and $64^3$ in equilibrium (dashed lines,
      averaged over 1431 initial random-field configurations). The inset shows
      the scaling collapse of curves in the main panel, using the same set of
      exponents for both equilibrium and non-equilibrium:
      $\tilde{d_\mm{a}}=0.81$, $\tau+\sigma\beta\delta=2.01$.}  
  \end{figure}

  Third, we measure the field integrated avalanche surface area distribution
  for given sizes at $R_\mm{c}$ in both equilibrium and non-equilibrium. Analogous to the
  avalanche time distribution obtained in 
  the non-equilibrium study~\cite{Perkovic-99}, the scaling form of surface
  area distribution can be written as: \be D_\mm{a}^{\mm{(int)}} (S,a) \sim
  a^{-(\tau+\sigma\beta\delta+\tilde{d_\mm{a}})/\tilde{d_\mm{a}}}   \
  \mc{D}_{\mm{a}}^{\mm{(int)}} (a/S^{\tilde{d_\mm{a}}}) \label{Daint} \ee  
  with $\tilde{d_\mm{a}} \equiv d_\mm{a}/d_\mm{S}$. Fig.~\ref{fig:Dint_a-S}
  shows the surface area distributions for different avalanche sizes and
  collapses of those curves using Eq.~\ref{Daint}. Here, we use
  $\tilde{d_\mm{a}}=0.81$ and $\tau+\sigma\beta\delta=2.01$ for both
  equilibrium and non-equilibrium collapses. The values are also consistent
  with what we obtained from the study of the field integrated avalanche size
  distribution and the fractal dimensions of avalanches. We find that with the
  same set of exponents, the scaling function $\mc{D}_{\mm{a}}^{\mm{(int)}}
  (X)$ in equilibrium and non-equilibrium match very well.

  In summary, we have shown that the equilibrium and non-equilibrium DIPTs of the
  zt-GRFIM behave surprisingly similarly in critical exponents, scaling
  functions and spatial structures of avalanches and clusters. Also, they 
  both obey the no-passing rule. All of these results indicate that the two
  DIPTs are very likely in the same universality class. Larger system sizes
  could be a direct way to test it further, especially for the fractal
  dimensions and anisotropy measures. Different disorder distributions and
  lattice types would also be useful methods to test the universality. Also,
  we want to emphasize the connection between all the known DIPTs associated
  with different dynamics (history dependence). As we know, the
  demagnetization curve displays a very similar DIPT as that of the saturation
  loop and the ground state~\cite{Carpenter-03, Colaiori-04}. Together with
  our new result, we suggests that all the three DIPTs, associated with the
  saturation loop, the demagnetization curve and the equilibrium magnetization
  curve respectively, are indeed in the same universality class. This would be
  very exciting. So far there is no RG treatment for the demagnetization curve,
  while there is for the saturation loop~\cite{Karin-96}. Motivated by these
  results, we find that analytic studies comparing the RG descriptions of the
  DIPTs with different dynamics are indeed an exciting prospect.

  We thank James P. Sethna, Andrew Dolgert, A. Alan Middleton, Y. Oono, J. Carpenter,
  R. White, M. Delgado, and G. Poore for valuable discussions. We acknowledge
  the support of NSF Grant No. DMR 03-14279 and NSF Grant No. DMR 03-25939 ITR
  (Materials Computation Center). This work was conducted on the Beowolf
  cluster of the Materials Computation Center at UIUC. 
  \bibliography{RFIM}

\begin{thebibliography}{25}
\expandafter\ifx\csname natexlab\endcsname\relax\def\natexlab#1{#1}\fi
\expandafter\ifx\csname bibnamefont\endcsname\relax
  \def\bibnamefont#1{#1}\fi
\expandafter\ifx\csname bibfnamefont\endcsname\relax
  \def\bibfnamefont#1{#1}\fi
\expandafter\ifx\csname citenamefont\endcsname\relax
  \def\citenamefont#1{#1}\fi
\expandafter\ifx\csname url\endcsname\relax
  \def\url#1{\texttt{#1}}\fi
\expandafter\ifx\csname urlprefix\endcsname\relax\def\urlprefix{URL }\fi
\providecommand{\bibinfo}[2]{#2}
\providecommand{\eprint}[2][]{\url{#2}}

\bibitem[{\citenamefont{Belanger and Nattermann}()}]{Nattermann-98}
\bibinfo{author}{\bibfnamefont{D.~P.} \bibnamefont{Belanger}} \bibnamefont{and}
  \bibinfo{author}{\bibfnamefont{T.}~\bibnamefont{Nattermann}},
  \bibinfo{note}{in \emph{Spin Glasses and Random Fields}, edited by A. P.
  Young (World Scientific, Singapore, 1998)}.

\bibitem[{\citenamefont{Villain}(1984)}]{Villain-84}
\bibinfo{author}{\bibfnamefont{J.}~\bibnamefont{Villain}},
  \bibinfo{journal}{Phys. Rev. Lett.} \textbf{\bibinfo{volume}{52}},
  \bibinfo{pages}{1543} (\bibinfo{year}{1984}).

\bibitem[{\citenamefont{Fisher}(1986)}]{Fisher-86}
\bibinfo{author}{\bibfnamefont{D.~S.} \bibnamefont{Fisher}},
  \bibinfo{journal}{Phys. Rev. Lett.} \textbf{\bibinfo{volume}{56}},
  \bibinfo{pages}{416} (\bibinfo{year}{1986}).

\bibitem[{\citenamefont{Ogielski}(1986)}]{Ogielski-86-b}
\bibinfo{author}{\bibfnamefont{A.~T.} \bibnamefont{Ogielski}},
  \bibinfo{journal}{Phys. Rev. Lett.} \textbf{\bibinfo{volume}{57}},
  \bibinfo{pages}{1251} (\bibinfo{year}{1986}).

\bibitem[{\citenamefont{Cherkassky and Goldberg}(1997)}]{Cherkassky-97}
\bibinfo{author}{\bibfnamefont{B.}~\bibnamefont{Cherkassky}} \bibnamefont{and}
  \bibinfo{author}{\bibfnamefont{A.~V.} \bibnamefont{Goldberg}},
  \bibinfo{journal}{Algorithmica} \textbf{\bibinfo{volume}{19}},
  \bibinfo{pages}{390} (\bibinfo{year}{1997}).

\bibitem[{\citenamefont{Sethna et~al.}(1993)\citenamefont{Sethna, Dahmen,
  Kartha, Krumhansl, Roberts, and Shore}}]{Sethna-93}
\bibinfo{author}{\bibfnamefont{J.~P.} \bibnamefont{Sethna}},
  \bibinfo{author}{\bibfnamefont{K.~A.} \bibnamefont{Dahmen}},
  \bibinfo{author}{\bibfnamefont{S.}~\bibnamefont{Kartha}},
  \bibinfo{author}{\bibfnamefont{J.~A.} \bibnamefont{Krumhansl}},
  \bibinfo{author}{\bibfnamefont{B.~W.} \bibnamefont{Roberts}},
  \bibnamefont{and} \bibinfo{author}{\bibfnamefont{J.~D.} \bibnamefont{Shore}},
  \bibinfo{journal}{Phys. Rev. Lett.} \textbf{\bibinfo{volume}{70}},
  \bibinfo{pages}{3347} (\bibinfo{year}{1993}).

\bibitem[{\citenamefont{Dahmen and Sethna}(1996)}]{Karin-96}
\bibinfo{author}{\bibfnamefont{K.~A.} \bibnamefont{Dahmen}} \bibnamefont{and}
  \bibinfo{author}{\bibfnamefont{J.~P.} \bibnamefont{Sethna}},
  \bibinfo{journal}{Phys. Rev. B} \textbf{\bibinfo{volume}{53}},
  \bibinfo{pages}{14872} (\bibinfo{year}{1996}).

\bibitem[{\citenamefont{Berger et~al.}(2000)\citenamefont{Berger, Inomata,
  Jiang, Pearson, and Bader}}]{Berger-00}
\bibinfo{author}{\bibfnamefont{A.}~\bibnamefont{Berger}},
  \bibinfo{author}{\bibfnamefont{A.}~\bibnamefont{Inomata}},
  \bibinfo{author}{\bibfnamefont{J.~S.} \bibnamefont{Jiang}},
  \bibinfo{author}{\bibfnamefont{J.~E.} \bibnamefont{Pearson}},
  \bibnamefont{and} \bibinfo{author}{\bibfnamefont{S.~D.} \bibnamefont{Bader}},
  \bibinfo{journal}{Phys. Rev. Lett.} \textbf{\bibinfo{volume}{85}},
  \bibinfo{pages}{4176} (\bibinfo{year}{2000}).

\bibitem[{\citenamefont{Schneider and Pytte}(1977)}]{Schneider-77}
\bibinfo{author}{\bibfnamefont{T.}~\bibnamefont{Schneider}} \bibnamefont{and}
  \bibinfo{author}{\bibfnamefont{E.}~\bibnamefont{Pytte}},
  \bibinfo{journal}{Phys. Rev. B} \textbf{\bibinfo{volume}{15}},
  \bibinfo{pages}{1519} (\bibinfo{year}{1977}).

\bibitem[{\citenamefont{Feldman}(2002)}]{Feldman-02}
\bibinfo{author}{\bibfnamefont{D.~E.} \bibnamefont{Feldman}},
  \bibinfo{journal}{Phys. Rev. Lett.} \textbf{\bibinfo{volume}{88}},
  \bibinfo{pages}{177202} (\bibinfo{year}{2002}), \bibinfo{note}{and references
  therein.}

\bibitem[{\citenamefont{Maritan et~al.}(1994)\citenamefont{Maritan, Cieplak,
  Swift, and Banavar}}]{Maritan-94}
\bibinfo{author}{\bibfnamefont{A.}~\bibnamefont{Maritan}},
  \bibinfo{author}{\bibfnamefont{M.}~\bibnamefont{Cieplak}},
  \bibinfo{author}{\bibfnamefont{M.~R.} \bibnamefont{Swift}}, \bibnamefont{and}
  \bibinfo{author}{\bibfnamefont{J.~R.} \bibnamefont{Banavar}},
  \bibinfo{journal}{Phys. Rev. Lett.} \textbf{\bibinfo{volume}{72}},
  \bibinfo{pages}{946} (\bibinfo{year}{1994}).

\bibitem[{\citenamefont{Perkovi\ifmmode~\acute{c}\else \'{c}\fi{}
  et~al.}(1999)\citenamefont{Perkovi\ifmmode~\acute{c}\else \'{c}\fi{}, Dahmen,
  and Sethna}}]{Perkovic-99}
\bibinfo{author}{\bibfnamefont{O.}~\bibnamefont{Perkovi\ifmmode~\acute{c}\else
  \'{c}\fi{}}}, \bibinfo{author}{\bibfnamefont{K.~A.} \bibnamefont{Dahmen}},
  \bibnamefont{and} \bibinfo{author}{\bibfnamefont{J.~P.}
  \bibnamefont{Sethna}}, \bibinfo{journal}{Phys. Rev. B}
  \textbf{\bibinfo{volume}{59}}, \bibinfo{pages}{6106} (\bibinfo{year}{1999}).

\bibitem[{\citenamefont{Hartmann}(2002)}]{Hartmann-02}
\bibinfo{author}{\bibfnamefont{A.~K.} \bibnamefont{Hartmann}},
  \bibinfo{journal}{Phys. Rev. B} \textbf{\bibinfo{volume}{65}},
  \bibinfo{pages}{174427} (\bibinfo{year}{2002}).

\bibitem[{\citenamefont{P\'erez-Reche and Vives}(2004)}]{Vives-04}
\bibinfo{author}{\bibfnamefont{F.~J.} \bibnamefont{P\'erez-Reche}}
  \bibnamefont{and} \bibinfo{author}{\bibfnamefont{E.}~\bibnamefont{Vives}},
  \bibinfo{journal}{Phys. Rev. B} \textbf{\bibinfo{volume}{70}},
  \bibinfo{pages}{214422} (\bibinfo{year}{2004}).

\bibitem[{\citenamefont{Colaiori et~al.}(2004)\citenamefont{Colaiori, Alava,
  Durin, Magni, and Zapperi}}]{Colaiori-04}
\bibinfo{author}{\bibfnamefont{F.}~\bibnamefont{Colaiori}},
  \bibinfo{author}{\bibfnamefont{M.~J.} \bibnamefont{Alava}},
  \bibinfo{author}{\bibfnamefont{G.}~\bibnamefont{Durin}},
  \bibinfo{author}{\bibfnamefont{A.}~\bibnamefont{Magni}}, \bibnamefont{and}
  \bibinfo{author}{\bibfnamefont{S.}~\bibnamefont{Zapperi}},
  \bibinfo{journal}{Phys. Rev. Lett.} \textbf{\bibinfo{volume}{92}},
  \bibinfo{pages}{257203} (\bibinfo{year}{2004}).

\bibitem[{\citenamefont{Carpenter and Dahmen}(2003)}]{Carpenter-03}
\bibinfo{author}{\bibfnamefont{J.~H.} \bibnamefont{Carpenter}}
  \bibnamefont{and} \bibinfo{author}{\bibfnamefont{K.~A.}
  \bibnamefont{Dahmen}}, \bibinfo{journal}{Phys. Rev. B}
  \textbf{\bibinfo{volume}{67}}, \bibinfo{eid}{020412} (\bibinfo{year}{2003}).

\bibitem[{\citenamefont{Sethna et~al.}(2001)\citenamefont{Sethna, Dahmen, and
  Myers}}]{Sethna-01}
\bibinfo{author}{\bibfnamefont{J.~P.} \bibnamefont{Sethna}},
  \bibinfo{author}{\bibfnamefont{K.~A.} \bibnamefont{Dahmen}},
  \bibnamefont{and} \bibinfo{author}{\bibfnamefont{C.~R.} \bibnamefont{Myers}},
  \bibinfo{journal}{Nature} \textbf{\bibinfo{volume}{410}},
  \bibinfo{pages}{242} (\bibinfo{year}{2001}).

\bibitem[{\citenamefont{Middleton and Fisher}(2002)}]{Middleton-02}
\bibinfo{author}{\bibfnamefont{A.~A.} \bibnamefont{Middleton}}
  \bibnamefont{and} \bibinfo{author}{\bibfnamefont{D.~S.}
  \bibnamefont{Fisher}}, \bibinfo{journal}{Phys. Rev. B}
  \textbf{\bibinfo{volume}{65}}, \bibinfo{pages}{134411}
  (\bibinfo{year}{2002}).

\bibitem[{\citenamefont{Hartmann}(1998)}]{Hartmann-98}
\bibinfo{author}{\bibfnamefont{A.~K.} \bibnamefont{Hartmann}},
  \bibinfo{journal}{PHYSICA A} \textbf{\bibinfo{volume}{248}},
  \bibinfo{pages}{1} (\bibinfo{year}{1998}).

\bibitem[{\citenamefont{Frontera et~al.}(2000)\citenamefont{Frontera,
  Goicoechea, Ort\'in, and Vives}}]{Vives-00}
\bibinfo{author}{\bibfnamefont{C.}~\bibnamefont{Frontera}},
  \bibinfo{author}{\bibfnamefont{J.}~\bibnamefont{Goicoechea}},
  \bibinfo{author}{\bibfnamefont{J.}~\bibnamefont{Ort\'in}}, \bibnamefont{and}
  \bibinfo{author}{\bibfnamefont{E.}~\bibnamefont{Vives}}, \bibinfo{journal}{J.
  Comp. Phys.} \textbf{\bibinfo{volume}{160}}, \bibinfo{pages}{117}
  (\bibinfo{year}{2000}).

\bibitem[{\citenamefont{Kuntz et~al.}(1999)\citenamefont{Kuntz, Perkovi\'{c},
  Dahmen, Roberts, and Sethna}}]{Kuntz-99}
\bibinfo{author}{\bibfnamefont{M.~C.} \bibnamefont{Kuntz}},
  \bibinfo{author}{\bibfnamefont{O.}~\bibnamefont{Perkovi\'{c}}},
  \bibinfo{author}{\bibfnamefont{K.~A.} \bibnamefont{Dahmen}},
  \bibinfo{author}{\bibfnamefont{B.~W.} \bibnamefont{Roberts}},
  \bibnamefont{and} \bibinfo{author}{\bibfnamefont{J.~P.}
  \bibnamefont{Sethna}}, \bibinfo{journal}{Comp. Sci. Eng.}
  \textbf{\bibinfo{volume}{1}}, \bibinfo{pages}{73} (\bibinfo{year}{1999}).

\bibitem[{\citenamefont{Middleton}(1992)}]{Middleton-92}
\bibinfo{author}{\bibfnamefont{A.~A.} \bibnamefont{Middleton}},
  \bibinfo{journal}{Phys. Rev. Lett.} \textbf{\bibinfo{volume}{68}},
  \bibinfo{pages}{670} (\bibinfo{year}{1992}).

\bibitem[{\citenamefont{Liu and Dahmen}(2006)}]{Liu-06}
\bibinfo{author}{\bibfnamefont{Y.}~\bibnamefont{Liu}} \bibnamefont{and}
  \bibinfo{author}{\bibfnamefont{K.~A.} \bibnamefont{Dahmen}},
  \bibinfo{journal}{to be published}  (\bibinfo{year}{2006}).

\bibitem[{\citenamefont{Frontera and Vives}(2002)}]{Vives-02}
\bibinfo{author}{\bibfnamefont{C.}~\bibnamefont{Frontera}} \bibnamefont{and}
  \bibinfo{author}{\bibfnamefont{E.}~\bibnamefont{Vives}},
  \bibinfo{journal}{Comp.\ Phys.\ Comm.} \textbf{\bibinfo{volume}{147}},
  \bibinfo{pages}{455} (\bibinfo{year}{2002}).

\bibitem[{\citenamefont{Jagodzinski et~al.}(1992)\citenamefont{Jagodzinski,
  Eisenriegler, and Kremer}}]{Jagodzinski-92}
\bibinfo{author}{\bibfnamefont{O.}~\bibnamefont{Jagodzinski}},
  \bibinfo{author}{\bibfnamefont{E.}~\bibnamefont{Eisenriegler}},
  \bibnamefont{and} \bibinfo{author}{\bibfnamefont{K.}~\bibnamefont{Kremer}},
  \bibinfo{journal}{J. Phys. I France} \textbf{\bibinfo{volume}{2}},
  \bibinfo{pages}{2243} (\bibinfo{year}{1992}).

\end{thebibliography}
\end{document}